\documentclass[floatfix,showpacs,aps,onecolumn,jkps,preprint,showkeys]{revtex4-1}
\usepackage{amssymb}
\pdfoutput=1

\usepackage[dvips]{epsfig}
\usepackage[english]{babel}
\usepackage{verbatim}
\usepackage{bbm}
\usepackage{caption}
\usepackage{array}
\usepackage[T1]{fontenc}
\usepackage{amsmath}
\usepackage{xcolor}
\usepackage{hyperref}
\hypersetup{colorlinks=true,linkbordercolor=blue,linkcolor=blue, citecolor=blue}
\usepackage{bm}
\usepackage{latexsym}
\usepackage{epsfig}
\usepackage{pbox}
\usepackage{multirow}
\usepackage{rotating}
\usepackage{color}
\usepackage{mathrsfs}
\usepackage{amsmath,amsfonts,amssymb,textcomp}
\usepackage{placeins}
\usepackage{xcolor}
\usepackage{hyperref}

\usepackage{subeqnarray}

\usepackage{epstopdf}
\usepackage{tensor}
\usepackage{indentfirst} 

\makeatletter
\renewcommand*\l@section{\@dottedtocline{1}{1.5em}{2.3em}}
\makeatother
\RequirePackage[normalem]{ulem} 
\RequirePackage{color}\definecolor{RED}{rgb}{1,0,0}\definecolor{BLUE}{rgb}{0,0,1} 

\usepackage{placeins}
\usepackage{float}

\begin{document}
\title{Gravitational entropy of wormholes with exotic matter and in galactic halos}

\author{R. de C. Lima$^{1}$}\email{rc.lima@unesp.br}
\author{Jos\'e A. C. Nogales$^{2}$}\email{jnogales@dfi.ufla.br}
\author{S. H. Pereira$^{3}$}\email{s.pereira@unesp.br}

\affiliation{\\${^{1,3}}$Universidade Estadual Paulista (Unesp)\\Faculdade de Engenharia, Guaratinguet\'a, Departamento de F\'isica e Qu\'imica\\
12516-410, Guaratinguet\'a, SP, Brazil.\\ ${^{2}}$Departamento de F\'isica,
Universidade Federal de Lavras (UFLA),
Caixa Postal 3037, 37200-000 Lavras, MG, Brazil.}

\begin{abstract}
\vspace{0.15cm}
\textcolor{black}{\hrule}
\vspace{0.3cm}
In this
work we study and compare the features of gravitational entropy near the throat of transversable wormholes formed by exotic matter
and wormholes in galactic halos. We have verified that gravitational entropy and entropy density 
of these wormholes in regions near their throats are indistinguishable for objects of same throat, 
despite the fact they are described by different metrics and by distinct energy-momentum tensors. We have found 
that the gravitational entropy density diverges near the throat for both cases, probably due to a non-trivial topology at this point, however allowing the interesting interpretation that a maximum flux of information can be carried through the throat of these wormholes.
In addition, we have found that both are endowed with an entropic behaviour similar to Hawking-Bekenstein's
entropy of non-rotating and null charge black holes.
\vspace{0.15cm}
\textcolor{black}{\hrule}
\vspace{0.8cm} 
\end{abstract}

\keywords{Wormholes; gravitational entropy; galactic halos}

\pacs{04.20.-q}

\maketitle

\section{Introduction}
\indent 
\indent Recently the number of works studying physical properties of the relativistic object called wormhole has increased in the literature. Several studies of wormholes in various cosmological and astrophysical scenarios have appeared in last years, including its existence in galactic halos \cite{1}, its gravitational lensing effect \cite{2,jusufi}, 
neutron star interiors \cite{3}, quark matter wormholes \cite{4}, solutions in Einstein-Cartan gravity \cite{reza} and in $R^2$-gravity \cite{sahoo}, and also the possibility that super-massive black holes  at the center of galaxies might be wormholes created in the early universe, connecting either two different
regions of our universe or two different universes in a multiverse model \cite{5}.

Predicted for the first time in 1916 by L. Flamm \cite{9} and
discussed later in a geometrical perspective by Einstein and Rosen in 1935 \cite{10}, a peculiar cha\-rac\-te\-ris\-tic appears for such objects connecting two four-dimensional manifolds, which was soon named 
Einstein-Rosen bridge (ER). In the fifties it was renamed by John Wheeler as wormholes, gaining prominence not only as a peculiar object of science fiction but also as a relativistic object of great phy\-si\-cal interest by renowned scientists \cite{11, 12}. In 1988, Kip Thorne and Michael S. Morris studied the e\-ner\-gy and topological 
conditions for a wormhole to be transversable \cite{12}. As a result they found that the region of the throat should contains something that creates a negative
pressure and violates the null energy condition \cite{visser98}. This kind of exotic matter capable of maintain the stability of the wormhole could be represented by scalar fields with negative kinetic energy for instance \cite{kashar}.\\
\indent This work is motivated by the implementation of the Penrose's conjecture extended by Rudjord and Gron, an intrinsic 
entropy associated to space-time \cite{6}-\cite{7}, along with the recent studies performed by Kuhffitig \cite{1}-\cite{2},
which in turn checks the possibility of formation of these objects of non-trivial to\-po\-lo\-gy in regions of galactic halos. We study
the behaviour of gravitational entropy evolution associated with these objects in the presence of exotic matter and in galactic 
halos. We have obtained that the total gravitational entropy and entropy density are indistinguishable for wormholes of exotic matter
and in galactic halos for wormholes of same throat, even though they are originated from 
distinct energy-momentum tensors and characterized by different boundary conditions and form functions. Also, we have found that the entropy
density of a wormhole in presence of exotic matter differs from that obtained in \cite{8}, once they use the same prescription as \cite{7}, where the spatial components used for the description of entropy and gravitational entropy density differ precisely to ensure a convergent and finite entropy density in the throat of these objects.  Already, in our results, these components are similar, leading to a divergent behavior near the throat of the wormhole, which corroborates with the non-trivial topology at this point, and decreasing asymptotically far
from the throat. This allows the interesting physical interpretation that a maximum flux of information associated with the spacetime of these wormholes occurs and can be carried through the throat. In addition, for both cases we have found
that the gravitational entropy of the wormholes are indistinguishable from the entropy 
of Hawking-Bekenstein on the surface
of non-rotating and null charge black holes, once the entropy is also proportional to the surface area. 

The paper is organized as follows. Section II presents the main equations concerning the definition of Rudjord
and Gron for the gravitational entropy, Section III presents the explicit results for the entropy of a wormhole in presence of
exotic matter and of the entropy in galactic halo. The conclusions are presented in Section IV. 

\section{Gravitational entropy}
\indent 
\indent Roger Penrose conjectured that the space-time information emerges from the microstates configurations of spacetime to 
the most fundamental level, establishing an analogy with the entropy of Boltzmann, which lists the macro and micro-states of 
matter \cite{6}. Since there is no well-defined quantum theory of gravitation, Penrose only evaluated the macroscopic 
behavior of spacetime, seeking for possible covariant candidates that could represent an intrinsic entropy of spacetime. 

The Einstein's equation in the form ${G_{\mu\nu}= R_{\mu\nu} - \frac{1}{2}g_{\mu\nu}R = -\kappa T_{\mu\nu}}$ shows that the Ricci tensor components as well as 
their invariants become null in a region of space-time that does not contain matter/energy, since the scalar curvature is null in this case. Faced with this scenario, Penrose realized 
that the Weyl tensor is nonzero even in the absence of energy, implying that this tensor must contains intrinsic 
information of spacetime \cite{13}. The Weyl tensor can be expressed in four dimensions by:
\begin{equation}\label{WEYLP}
 C_{\alpha\beta\gamma\delta} = R_{\alpha\beta\gamma\delta} - g_{\alpha[\gamma}R_{\delta]\beta} + g_{\beta[\gamma}R_{\delta]\alpha} + \frac{1}{3}Rg_{\alpha[\gamma}g_{\delta]\beta},
\end{equation}
where the brackets indicate a antisymmetric combination, ${A_{[\mu\nu]} = \frac{1}{2}(A_{\mu\nu}}$ ${- A_{\nu\mu})}$.
${R_{\alpha\beta\gamma\delta}}$ is the Riemann's tensor, ${R_{\alpha\beta}}$ is the Ricci's tensor and ${R}$ is the scalar 
curvature. Thus, the Weyl invariant \cite{6}, defined as
\begin{equation}
 W \equiv C^{\alpha\beta\gamma\delta}C_{\alpha\beta\gamma\delta},\label{IWEYL}
\end{equation}
must be proportional to the gravitational entropy of a black-hole.

Based on this hypothesis, Rudjord and Gron developed a work on the gravitational entropy in Schwarzschild black holes 
and showed that just using the Weyl tensor the gravitational entropy disagreed with the results expected by Hawking-Bekenstein \cite{13,16}. It was 
necessary to redefine this entropy. In order to results be similar to the values es\-ta\-bli\-shed by Hawking and Bekenstein entropy
to black-holes, it was necessary that the entropy in question continued to be proportional to the Weyl invariant however inversely
proportional to the invariant Krestschmann, whose definition is
\begin{equation}\label{IKres}
K \equiv R^{\alpha\beta\gamma\delta}R_{\alpha\beta\gamma\delta}.
\end{equation}
Thus the gravitational entropy must be given by the surface integral:
\begin{eqnarray}
S = k_s \int_\sigma {\bf \Psi}\cdot {\bf d\sigma},\quad {\bf \Psi} = P\hat{{\bf e_{r}}}, \label{S}
\end{eqnarray}
where $k_s=k_Bc^3/4G\hbar$ and ${{\bf \Psi}}$ represents the scalar ${P}$ oriented radially. With the square of this scale given by 
${P^{2}}$ = ${W/K}$ =
${C^{\alpha\beta\gamma\delta}C_{\alpha\beta\gamma\delta}(R^{\alpha\beta\gamma\delta}R_{\alpha\beta\gamma\delta})^{-1}}$, Rudjord
and Gron proposed that in order to obtain the gravitational entropy of a black hole with the same Schwarzschild results
obtained by Hawking and Bekenstein \cite{7}, the gravitational entropy should be defined as: 
\begin{equation}\label{um}
 S = k_{s}\int_{\sigma}\sqrt{C^{\alpha\beta\gamma\delta}C_{\alpha\beta\gamma\delta}(R^{\alpha\beta\gamma\delta}R_{\alpha\beta\gamma\delta})^{-1}}\hat{{\bf e_{r}}}\cdot \bf{d\sigma}.
\end{equation}
In the definition of $S$, the
area element under a spherical surface is given by
\begin{equation}\label{dois}
 {\bf d\sigma} = \frac{\sqrt{h}}{\sqrt{h_{rr}}}d\theta d\varphi\hat{{\bf e_{r}}},
\end{equation}
where ${h}$ is the determinant of spatial metric ${h_{ij}}$, of a spherically symmetric line element 
\begin{equation}\label{tres}
 h_{ij} = g_{ij} - \frac{g_{i0}g_{j0}}{g_{00}}.
\end{equation}

Using the Gauss's divergence theorem in (\ref{um}),
\begin{equation}
 \int_{\sigma}{\bf \Psi}\ldotp {\bf d\sigma} = \int_{V}(\nabla.{\bf\Psi})dV,
\end{equation}
we obtain the entropy density as
\begin{eqnarray}\label{sdef}
 s &=& k_{s} \vert\nabla\ldotp{\bf\Psi}\vert \nonumber\\
 &=&  k_{s}\vert\nabla\ldotp\sqrt{C^{\alpha\beta\gamma\delta}C_{\alpha\beta\gamma\delta}(R^{\alpha\beta\gamma\delta}R_{\alpha\beta\gamma\delta})^{-1}}\hat{{\bf e_{r}}}\vert.
\end{eqnarray}

\section{Boundary conditions and entropy gravitational for wormholes}
\indent 
\indent This section discusses under what physical and geometric conditions a wormhole can emerge in a certain region of space
from the existence of exotic matter and in galactic halos. Starting from the definition of the gravitational entropy already 
presented, we calculate the gravitational entropy and entropy density for these two cases. 

 
\subsection{Entropy of a wormhole in presence of exotic matter}
\indent 
\indent A metric in a static spherically symmetric space-time for a wormhole \cite{12} is described by
\begin{equation}\label{morris}
 ds^{2} = -e^{2\Phi(r)}dt^{2} + \frac{1}{1 - \frac{b(r)}{r}}dr^{2} + r^{2}(d\theta^{2} + \sin^{2}\theta d\varphi^{2}),
 \end{equation}
where $b(r)$ is the shape function satisfying $\lim_{r\to \infty}[b(r)/r]\to 0$ and ${\Phi(r)}$ is the redshift function satisfying $\lim_{r\to \infty}\Phi(r)\to 0$. The radial coordinate $r$ has a specific geometric 
meaning, where ${2\pi r}$ is the circumference of a circle centered on the throat of the wormhole. Thus, $r$ defines the 
ge\-o\-me\-tric limits of coverage of wormhole space, de\-cre\-a\-sing from $r={+ \infty}$ to a minimum value such that
${r_{0} = b_{0} = b(r_{0})}$ in the throat's radius, then in\-cre\-a\-sing to ${+\infty}$.

In his classic work, Morris and Thorne delimited a possible boundary condition to confine the exotic matter inside the wormhole, which will keep
the throat of a wormhole open \cite{12}. The complete solution for shape function  and redshift function in this case are:
\begin{equation}\label{bb}
 b(r) = \left\{\begin{array}{rl}
                (b_{0}r)^{\frac{1}{2}}, &   b_{0} \leq r \leq r_{c}\\
                \frac{1}{100r}, &  r_{c} \leq r \leq R_{s}\\
                \frac{1}{3}[\frac{(r^{3} - R_{s}^{3})}{R_{s}^{2}}] + \frac{R_{s}}{100}, &  R_{s} \leq r \leq R_{1} \\
                B = \frac{R_{s}}{100}, &   R_{1} \leq r
               \end{array}\right.
\end{equation}
and
\begin{equation}\label{fifi}
 \Phi(r) = \left\{\begin{array}{rl}
                \Phi_{0} \cong - 0,01, &  b_{0} \leq r \leq R_{1}\\
                \frac{1}{2}ln(1 - \frac{B}{r}), &  R_{1} \leq r, 
               \end{array}\right.
\end{equation}
where ${r_{c}} = 10^{4}b_{0}$, ${R_{1} = R_{s} + \Delta R}$ 
and ${\Delta R = \frac{1}{100R}}$. 
 
We are interested in the region close to the throat of the wormhole, namely $b_{0} \leq r \leq r_{c}$, thus we have the line element of this simple case \cite{8} described by:
\begin{eqnarray}\label{morrisi}
 ds^{2} = \chi dt^{2} + \frac{1}{1 - \sqrt{\frac{b_{0}}{r}}}dr^{2} + r^{2}(d\theta^{2} + \sin^{2}\theta d\varphi^{2}),\nonumber\\
 \end{eqnarray}
where ${\chi = -e^{2\Phi_{0}}}$, ${\Phi_{0} \cong -0,01}$ and ${b_{0} = r_{0}}$ is the radius of the throat.

Now we will determine the entropy of the system. For this we obtain the 
invariants of Weyl and Krestschmann for the metric \eqref{morrisi} using (\ref{IWEYL}) and (\ref{IKres}):
\begin{equation}\label{wi}
W_{1} = (C^{\alpha\beta\gamma\delta}C_{\alpha\beta\gamma\delta})_{1} = \frac{25b_{0}}{12r^5},
\end{equation}
\begin{equation}\label{ki}
K_{1} = (R^{\alpha\beta\gamma\delta}R_{\alpha\beta\gamma\delta})_{1} = \frac{9b_{0}}{2r^5}.
\end{equation} 
Thus, the scalar $P$ defined in terms of (\ref{wi}) and (\ref{ki}) as established by Rudjord-Gron for this case is
\begin{equation}
P_{1} = \sqrt{W_{1}/K_{1}} = {5\sqrt{6}\over 18}.\label{P1}
\end{equation} 
The infinitesimal surface element of the above metric can be obtained using the
definition (\ref{dois}) and (\ref{tres}):
\begin{eqnarray}
 {\bf d\sigma} = \frac{\sqrt{h}}{\sqrt{h_{rr}}}d\theta d\phi = r^{2}sin(\theta)d\theta d\phi\hat{{\bf e_{r}}}.
\end{eqnarray}

Using the scalar ${P_{1}}$ obtained for this situation, together with the surface element area found above, this region will have an
entropy given by
\begin{eqnarray}\label{Sent}
 S_{1} &=& k_{w}\int_{\sigma}{\bf\Psi}\cdot d{\bf\sigma}\nonumber\\
&=& k_{w}[P_{1}r^2 - P_{1}(\varepsilon)\varepsilon^{2}]\int_{0}^{2\pi}\int_{\theta_{0}}^{\theta} \sin(\theta)d\theta d\phi,\nonumber\\
\end{eqnarray}
where ${k_{w}}$ is a constant similar to $k_s$ in \eqref{S} that relates the entropy to the surface integral, and ${\varepsilon}$ a very small opening radius inside.

In order to ensure that the beam is close to the throat, and therefore approximately constant, we limit ${\theta}$ into regions where ${r_{0} \leq r \leq 2r_{0}}$. Taking the limit of ${S_{1}}$ when ${\varepsilon \rightarrow 0}$,
we have
\begin{eqnarray}\label{Si}
\lim_{\varepsilon \rightarrow 0}S_{1} & = & k_{w}\frac{5\sqrt{6}}{18}r^{2}\int_{0}^{2\pi}\int_{\theta_{0}}^{\theta}\sin(\theta)d\theta d\phi \nonumber\\ 
& = &(2\pi k_{w}(\cos(\theta_{0})-\cos(\theta)))\frac{5\sqrt{6}}{18}r^{2}.
 \end{eqnarray}\\
Defining $\zeta_{1}\equiv {2k_{w}(cos(\theta_{0})-cos(\theta))}$, we notice that ${S_{1}}$ is proportional 
to the area near the throat of the wormhole:
\begin{equation}\label{S1}
 \frac{S_{1}}{\zeta_{1}} = \frac{5\sqrt{6}}{18}\pi r^{2}.
\end{equation}

Using (\ref{sdef}) and the divergence operator, in a riemannian manifold, ac\-ting on a vector in curved spacetime, $\nabla\cdot {\bf \Psi}=(1/\sqrt{-g})\partial_i(\sqrt{-g}\,\Psi{^i})$, we obtain the entropy density (see \ref{AAA}): 
\begin{eqnarray}\nonumber
s_{1} = k_{w} \arrowvert {\bf \nabla}\ldotp{\bf \Psi} \arrowvert &=& k_{w} \left|\frac{1}{\sqrt{-g}}\frac{\partial}{\partial r}(\sqrt{-g}P_{1}(r))\right|,\nonumber
\end{eqnarray}
resulting in:
\begin{equation}\label{ss}
 s_{1} = k_{w}\left| -\frac{5}{72}\frac{\sqrt{6}\left(-9b_{0}+8r\sqrt{\frac{b_{0}}{r}}\right)}{\left(\sqrt{\frac{b_{0}}{r}}-1\right)r^{2}\sqrt{\frac{b_{0}}{r}}}\right|,
\end{equation}
where we have used (\ref{P1}) and ${\sqrt{-g} = \sqrt{-det(h_{ij})}}$, whose value is expressed as
\begin{eqnarray}\label{gdef}
\sqrt{-g}  =  r^{2}\sin(\theta)[(b_{0}/r)^{1/2}-1]^{-1/2},\nonumber
\end{eqnarray}
with ${\theta\neq n\pi}$.\\

\begin{figure}[!h]\centering
 \includegraphics[height=6.7cm]{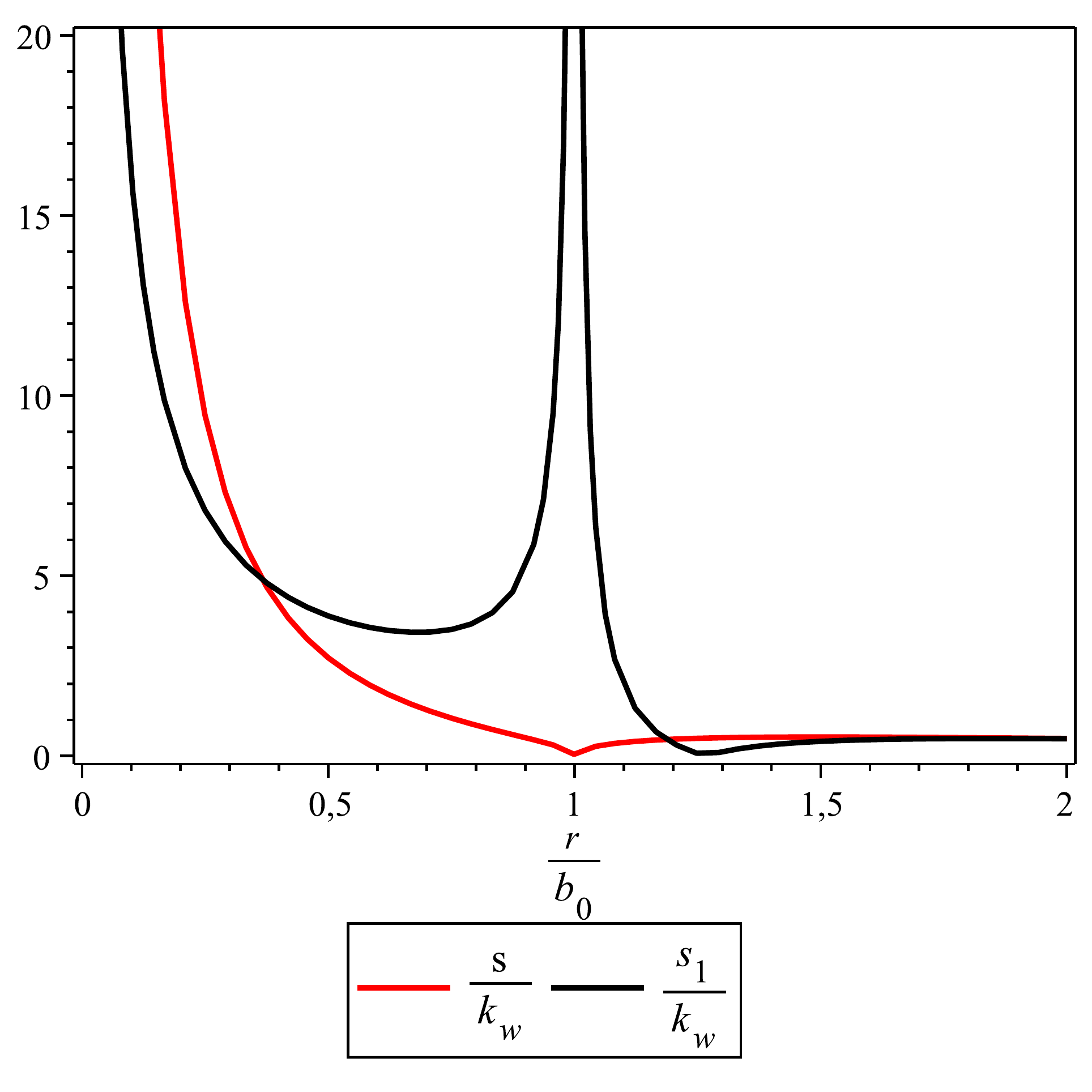}
 \caption{Comparison between gravitational entropy density of wormholes supported by exotic matter
 ${s/k_w}$ (red line) \cite{8} and ${s_{1}/k_w}$ (black line) for us.}\label{C0}
 \end{figure}

It is important to emphasize that this result for the gravitational entropy density in
wormholes with exotic matter is quite different from the result found in \cite{8}, where they use the spatial components of the vector ${{\bf\Psi}}$ like ${\Psi^{i} = (P/\sqrt{h_{rr}},0,0)}$, given by\cite{7}:
\begin{equation}\label{romeroetal}
 s = k_{w}\left|\sqrt{\frac{2}{27}}5\sqrt{\left[1-\sqrt{\frac{b_{0}}{r}}\right]}\frac{1}{r}\right|.
\end{equation}
The difference between the behaviors generated by \eqref{ss} and \eqref{romeroetal} is shown in the Fig.\ref{C0}.

We believe that the divergence of gravitational entropy density in ${r\longrightarrow b_{0}}$ or ${r_{th}}$, studied in this context, occurs probably due to structure of the synthesis between entropy density of space-time and gravitational dynamics, revealing that the 
spacetime has a non-trivial topology at this point. This suggests that the local entropy flux density, associated with the  Noether current for gravity\cite{Padmanabhan:2009zz}, must be analyzed near the wormhole's throat, since the information contained there must be indeterminate at ${r = b_{0}}$, and with no physical meaning at ${r< b_{0}}$. Moreover, from a physical point of view we know that entropy is associated with information and therefore,
the increasing gravitational entropy density near the throat of these wormholes is an indication that the flux of information associated with spacetime increasingly as it approaches the throat on the right.

\subsection{Entropy of a wormhole in galactic halo}
\indent 
\indent Recent studies \cite{1,2} have shown that regions in galactic halo ensure
necessary properties to support transversable wormholes, based on density profile 
of Navarro-Frenk-White \cite{14,15} to galaxies and the method for calculating the deflection angles of galactic halo.

The metric of such astrophysical system is 
described as
\begin{equation}\label{m1}
 ds^{2} = - Br^{l} dt^{2} + e^{2g(r)} dr^{2} + r^{2} ( d\theta^{2} + \sin^{2} \theta d\varphi^{2} ) ,
\end{equation}
where ${g(r)}$ is a function that defines the radial behaviour of the halo, ${B = 1/r_{s}^{l}}$, where
${r_{s}}$ is the ef\-fec\-ti\-ve radius of the galaxy, 
$l = 2v_{\phi}^{2}$ and ${v_{\phi}}$ is the rotational speed of galactic halo. In
typical galaxies, ${v_{\phi} \sim 10^{-3} c= 300km/s}$. 

Considering the Morris-Thorne's metric (\ref{morris}) and the 
metric for galactic halo (\ref{m1}), a wormhole supported by a galactic halo \cite{1}
have the line element in the form: 
\begin{equation}\label{mhalo}
ds^{2} = - \left(\frac{r}{r_{s}}\right)^{l}dt^{2} + \frac{1}{1-\frac{b(r)}{r}}dr^{2} + r^{2}(d\theta^{2} + sin^{2}\theta d\varphi^{2}).
\end{equation}

Through the $G_{00}$ Einstein field equations, the shape function $b(r)$ can be written as
\begin{equation}\label{db}
\frac{ b^{'}(r)}{r^2} = 8\pi\rho(r),
\end{equation}
where a prime denotes derivative with respect to ${r}$. As we are dealing with this wormhole in a halo, the energy density stemmed by this, according to Navarro-Frenk-White 
model \cite{14,15} is defined by
\begin{equation}\label{navarro}
 \rho(r) = \frac{\rho_{s}}{\frac{r}{r_{s}}\left(1 + \frac{r}{r_{s}}\right)^{2}},
\end{equation}
where ${\rho_{s}}$ is the effective density of the galaxy and ${r_{s}}$ is the effective radius. Replacing (\ref{navarro}) into
(\ref{db}), the shape function is:
\begin{eqnarray}\label{b1}
 b(r) &=& 8\pi\int\frac{\rho_{s}}{\frac{r}{r_{s}}\left(1 + \frac{r}{r_{s}}\right)^{2}}r^{2}dr\nonumber\\ 
      &=& 8\pi\rho_{s}r_{s}^{3}\left[ln\left(1 + \frac{r}{r_{s}}\right) + \frac{1}{1 + \frac{r}{r_{s}}} + C\right],\nonumber\\
\end{eqnarray}
where ${C}$ is a constant of integration. Defining a constant $\bar{r}={8\pi\rho_{s}r_{s}^{3}}$, is possible to analyse the conditions that guarantees a
transversable wormholes \cite{1,12}:

\begin{enumerate}
\item The redshift function ${\Phi(r)}$ prevents an event horizon;

\item For the shape function ${b(r)}$, it is valid that ${b(r_{th})}$ = ${r_{th}}$ and ${b^{'}(r_{th}) < 1}$, with ${r_{th}}$ named throat's radius;

\item For ${b(r)}$, we have ${b(r) < r}$ when ${r > r_{th}}$.
\end{enumerate}

Since ${r_{s}\gg r_{th}}$, the condition 2 leads to $b(r_{th}) = \bar{r} = r_{th}$ if we set $C=0$ into (\ref{b1}).

In his work, Kuhfittig \cite{2} made an analysis on the metric to verify the existence of
wormholes in halos by Bozz method, obtaining the values of the parameters as: Effective radius of the galaxy, ${r_{s} = 20,88 kpc = 6,44\times 10^{20}m}$; 
throat of the wormhole in this region, ${r_{th} = \bar{r} = 3,7466\times 10^{15}m}$ and parameter $l$ associated
with the typical rotation velocity in the galactic halo (for ${c=1}$) is ${l =2\times 10^{-6}}$. Replacing the shape
function of (\ref{b1}) the wormhole's metric in the galactic halo is:
\begin{equation}\label{mmhalo}
 ds^{2} = - \left(\frac{r}{r_{s}}\right)^{l}dt^{2} + \frac{dr^{2}}{1 - \frac{\bar{r}\ln\left(1 + \frac{r}{r_{s}}\right)}{r} - \frac{\bar{r}}{r\left(1 + \frac{r}{r_{s}}\right)}} + r^{2}d\Omega^{2},
\end{equation}
where ${d\Omega^{2} = d\theta^{2} + \sin^{2}\theta d\varphi^{2}}$.

Now, from (\ref{mmhalo}) we can determine the Weyl and  Kretschmann invariants (see Appendix),
and consequently the scalar $P$ for this case, which we call ${P_{2}(r)=\sqrt{W_{2}(r)/K_{2}(r)}}$. Similarly to the entropy 
calculation in the regions near of the throat, we choose ${r_{th} < r < 6r_{th}}$ and we obtain: 
\begin{eqnarray}
 S_{2} &=& k_{w}[P_{2}(r)r^2-P_{2}(\varepsilon)\varepsilon^2]\int_{0}^{2\pi}\int_{\theta_{0}}^{\theta}sin(\theta)d\theta d\phi \,,\nonumber\\
\end{eqnarray}
where ${k_{w}}$ is a constant. In the limit $\varepsilon \to 0$ we have
\begin{eqnarray}
  \lim_{\varepsilon\to 0}S_2     &=& 2\pi(\cos(\theta_{0})-\cos(\theta))k_{w}P_{2}(r)r^{2},\label{Sent1}
\end{eqnarray}
The interval
${[\theta_{0},\theta]}$ delimits the region for ${r}$ values near ${r_{th}}$. Defining ${\zeta_{2}\equiv 2 k_{w}(\cos(\theta_{0})-\cos(\theta))}$, 
we can rewrite (\ref{Sent1}) as
\begin{equation}\label{Sent2}
 \frac{S_{2}}{\zeta_{2}} = \pi r^{2}P_{2}(r).
\end{equation}

The entropy density is obtained via (\ref{sdef}). In this case:
\begin{eqnarray}\label{s1}
 s_{2} &=& k_{w}\left|\frac{1}{\sqrt{-g}}\frac{\partial}{\partial r}(\sqrt{-g}P_{2}(r)) \right|\nonumber\\ 
   &=& k_{w}\left|\frac{1}{2}\frac{\partial}{\partial r}(\ln(-g))P_{2}(r) + \frac{\partial}{\partial r}P_{2}(r)\right|,
\end{eqnarray}
where $g$ is the determinant of Morris-Thorne's matrix,
\begin{eqnarray}\label{g1}
 g = det\left(g_{ij} - \frac{g_{i0}g_{j0}}{g_{00}}\right) = \frac{-r^{5}sin^{2}(\theta)}{b(r)-r},
\end{eqnarray}
and ${b(r)}$ will be obtained through (\ref{b1}) to ensure that this parameter has the information 
of the galactic halo, with ${C = 0}$. Using (\ref{g1}) in (\ref{s1}), we have
\begin{eqnarray}\label{s2}
 \frac{s_{2}}{k_{w}}&=&\left|\frac{1}{2}\left(\frac{5(b(r) -r)+r}{r(b(r)-r)}\right)P_{2}(r) + \frac{\partial}{\partial r}P_{2}(r)\right|.\nonumber\\
\end{eqnarray}

\subsection{Gravitational entropy and entropy density of wormholes with exotic matter and in galactic halo}
\begin{figure}[!h]\centering
 \includegraphics[height=6.7cm]{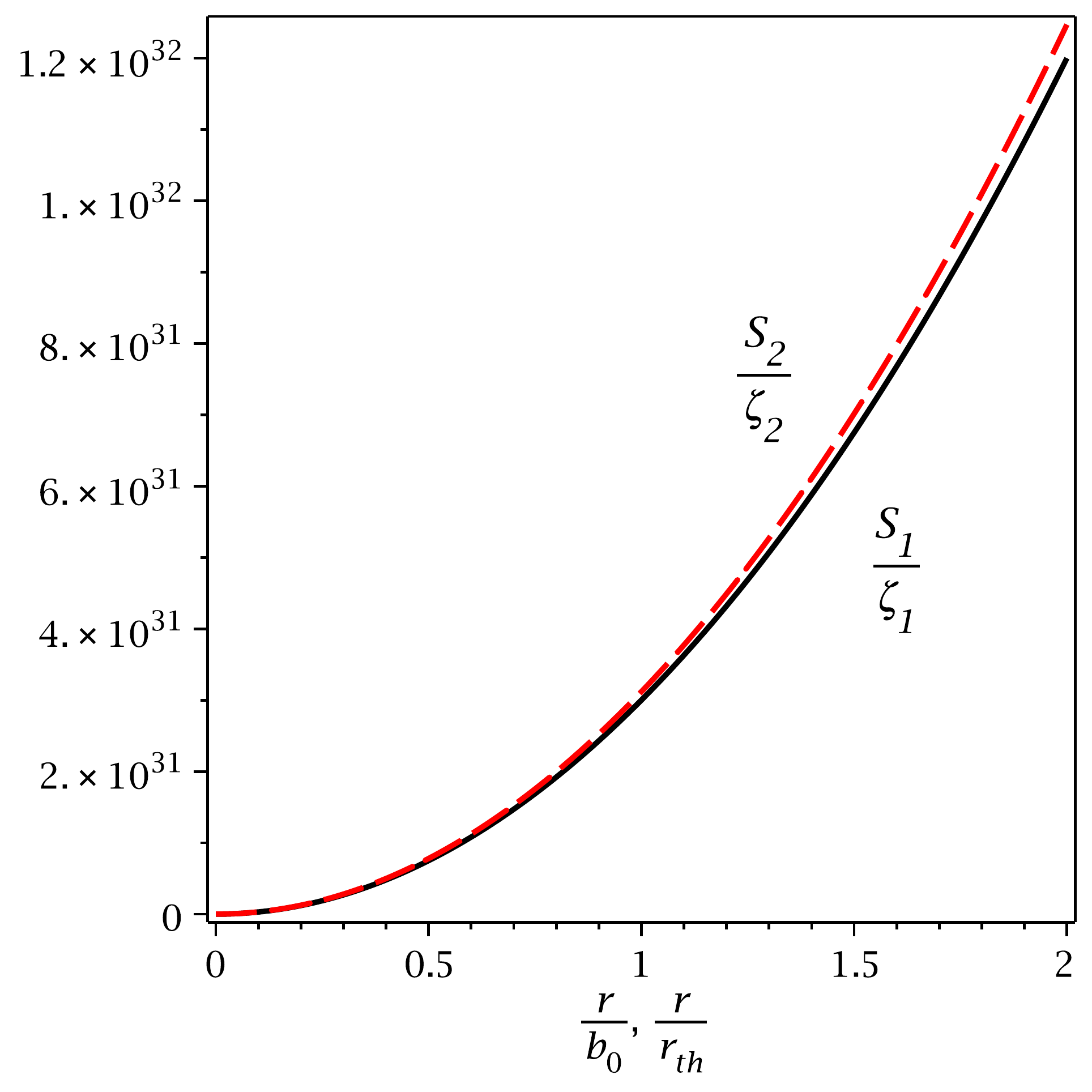}
 \caption{Comparison between gravitational entropy of wormholes supported by exotic matter
 ${S_{1}/\zeta_1}$ (black line) and in galactic halo ${S_{2}/\zeta_2}$ (red dashed line). The throats of these objects are indexed by
 ${b_{0}}$ and ${r_{th}}$, respectively, with $b_0=r_{th}=3,7466\times 10^{15}$m.}\label{C2}
 \end{figure}

\indent 
\indent Now we can compare the total gravitational entropy for wormholes with exotic matter and in galactic halo from equations
(\ref{S1}) and (\ref{Sent2}), respectively. The Fig.\ref{C2} shows the behaviour of $\frac{S_{1}}{\zeta_{1}}$ and
$\frac{S_{2}}{\zeta_{2}}$ as a function of $r/b_0$ for the wormhole with exotic matter and $r/r_{th}$ for the wormhole in the
galactic halo, with $b_0=r_{th}=3,7466\times 10^{15}$m. It easy to see that near the throats the gravitational entropy is
almost indistinguishable.  

The same behaviour is found when analysing the gravitational entropy densities obtained for wormholes
with exotic matter and in galactic halos,
des\-cri\-bed by the equations (\ref{ss}) and (\ref{s2}).
\begin{figure}[!h]\centering
 \includegraphics[height=6.7cm]{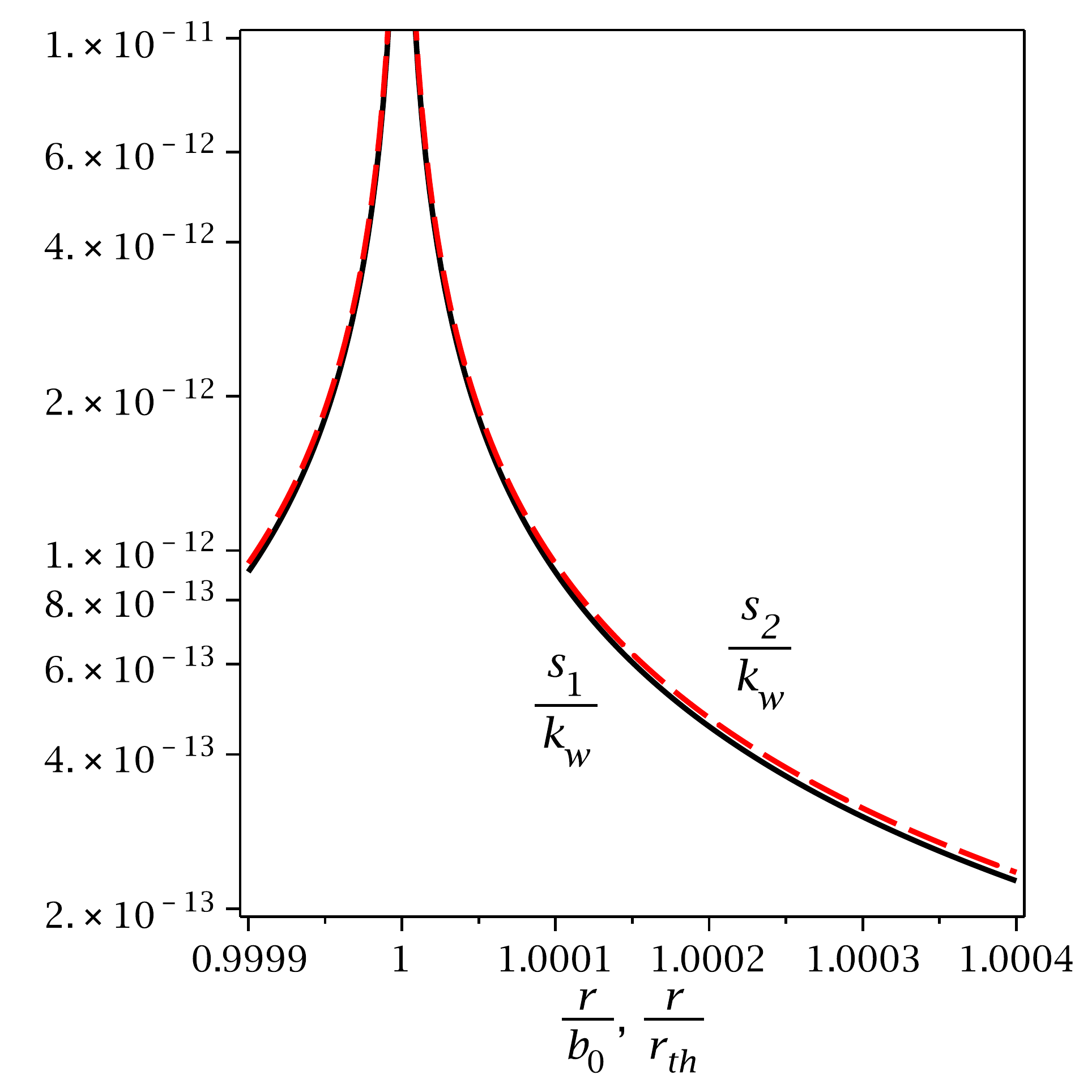}
 \caption{Comparison between gravitational entropy density of wormholes supported by exotic matter
 ${s_{1}/k_w}$ (black line) and in galactic halo ${s_{2}/k_w}$ (red dashed line), with $b_0=r_{th}=3,7466\times 10^{15}$m.}\label{C3}
 \end{figure}
The Fig.\ref{C3} shows that the entropy density diverges exactly in the throat for both cases and decrease as the radius increase, from ${b_{0}}$
and ${r_{th}}$. It is important to emphasize that for ${r < b_{0}}$ and ${r < r_{th}}$ the entropy density
has no physical meaning, since does not exist events in this region, because in this throat limit, the spacetime has 
non-trivial topology and any information associated with the manifold is lost.

\section{Concluding Remarks} 

In this work we have studied the behaviour of gravitational entropy near of the throat and for external radius in wormholes
emerging from exotic matter and in galactic halos. The main result obtained is that for both wormholes the total gravitational
entropy and the entropy density are indistinguishable for wormholes of same throat. 
This result is interesting because the first metric, \eqref{morris}, is a generic description of wormholes
whose source is a theoretical exotic matter, having boundary conditions that characterize the form function ${b(r)}$ given by expression
\eqref{bb}, while the second metric, \eqref{mmhalo}, has a source of energy-momentum from halos (dark matter) present in the
function ${b(r)}$ characterized by \eqref{b1}, completely different from the other one. Notice that the features of the expected
galaxies, in this framework, takes into account the radius and the effective density. Both have different sources and this does
not guarantee that they must have the same entropies and densities. Even with
Weyl and Kretschmann invariants represented by very different expressions, when calculating the scalar $P=\sqrt{W/K}$ the result for the Rudjord-Gron entropy is very similar. In addition, contrary to the results in \cite{8} for
the gravitational entropy density of wormholes in classic exotic matter, where the entropy density was shown to be null at the
throat, we have obtained that at this point it is divergent. The same occurs for wormholes in galactic halos. We interpret that this divergence occurs because, at the throat limit, the spacetime has non-trivial topology and thus the information contained must be indeterminate for ${r = b_{0}}$ or ${r = r_{th}}$. May be the circumstances behind this are associated with the Noether current for gravity and the local entropy flux density\cite{Padmanabhan:2009zz} for these objects in ${b_{0}}$, which must be better understood. On the other hand, from a physical point of view we know that entropy is associated with information and therefore,
the gravitational entropy density near the throat of these wormholes admits the interpretation that the flux of information associated with spacetime is high as it approaches the throat on the right, i.e., ${r\longrightarrow b_{0}^{+} \Rightarrow s(r)\longrightarrow s(b_{0}^{+})\cong s_{max}}$ with exotic matter [or ${r\longrightarrow r_{th}^{+} \Rightarrow s(r)\longrightarrow s(r_{th}^{+})\cong s_{max}}$ in galactic halos], contrary to the interpretation that the entropy flux (information) is null in ${b_{0}}$ by\cite{8}. It is reasonable to accept this interpretation that the flux is high in the throat once in this point there is the connection between two distinct regions of spacetime, concentrating as much information as possible.

Both wormholes have a similar behaviour with the entropy on the surface of non-rotating and null charge black holes, being proportional 
to its area, remembering that the entropy of Hawking-Bekenstein for these objects is 
proportional to the radius squared of the event horizon, as shown in \cite{16}
\begin{equation}\label{SBH}
 S = k.4\pi R^{2} = k.A = S_{HB},
\end{equation}
where ${k}$ is a constant relating entropy and area ${A}$ of the
event horizon of radius $R$. This shows that the entropy
of Hawking-Bekenstein for Schwarzschild's black holes and gravitational entropy for 
wormholes with exotic matter and in galactic halo are also indistinguishable, once the entropy is also proportional
to the surface area. So, reinforcing the work \cite{5}, even in the entropic aspect there is a subtle break when dealing
with the distinguishing of wormholes generated by exotic matter or in halos and black holes of Hawking-Bekenstein, as
verified by the results \eqref{S1} and \eqref{Sent2} compared to \eqref{SBH}. In other words, the space-time itself may account
for the entropic characteristics in this prescription, regardless of the content of the scenarios studied.

As already mentioned in \cite{5}, supermassive black holes at the
center of galaxies are good candidates to be wormholes. At the same time,
black hole thermodynamic is well studied theoretically. Faced with this situation, if the entropic behaviour of
wormholes are similar to black holes, we have strong motivations to studying their properties, aiming to predict
possible thermodynamic stability and astrophysical nature of such exotic objects. We hope that the results here 
discussed may serve also to motivate investigations about behaviors between other wormholes generated from different energy-matter, 
besides arouse studies of the correlation between entropy and spacetime, according to Penrose-Rudjord-Gron, as
well as to what extent sources of distinct wormholes can generate indistinguishable entropies or have behaviors similar to black holes.

\section*{Acknowledgments}
RdeCL is grateful to Coordena\c c\~ao de Aperfei\c coamento de Pessoal de N\'ivel Superior - Brasil (CAPES) - Finance Code 001 for financial support. SHP is grateful to CNPq - Conselho Nacional de 
Desenvolvimento Cient\'ifico e Tecnol\'ogico, Brazilian research agency, for financial support, grants numbers 303583/2018-5 and 400924/2016-1. 

\appendix

\section{Components of vector ${{\bf \Psi}}$ for entropy density}\label{AAA}

A vector it is described as ${{\bf v} = v^{\mu}{\bf \hat{e}_{\mu}} = v_{\mu}{\bf \hat{e}^{\mu}}}$, where $v^{\mu}$ (or $v_{\mu}$) is its components and ${{\bf \hat{e}_{\mu}}}$ (or ${\bf{\hat{e}^{\mu}}}$) is its basis. Admitting ${{\bf\Psi}}$ as a vector in curved spacetime and with radial symmetry (our interest) one obtain, from Eq. \eqref{S}:

\begin{eqnarray}\label{2AAA}
  S &=& k_{s}\int_{\sigma}{\bf\Psi}\cdot {\bf d\sigma} = k_{s}\int_{\sigma}\Psi^{\mu}{\bf \hat{e}_{\mu}}\cdot {\bf d\sigma} = k_{s}\int_{\sigma}(\Psi^{t}{\bf \hat{e}_{t}}+\Psi^{r}{\bf \hat{e}_{r}}+\Psi^{\theta}{\bf\hat{e}_{\theta}}+\Psi^{\varphi}{\bf\hat{e}_{\varphi}})\cdot {\bf d\sigma}\nonumber\\
   &=& k_{s}\int_{\sigma}\Psi^{r}{\bf \hat{e}_{r}}\cdot {\bf d\sigma} = k_{s}\int_{\sigma}P(r){\bf\hat{e}_{r}}\cdot {\bf d\sigma}, 
\end{eqnarray}
where ${{\bf\Psi} = P(r){\bf\hat{e}_{r}}}$, with ${P(r)\equiv\sqrt{W(r)/K(r)}}$, whose ${W(r)}$ and ${K(r)}$ are Weyl and Kretschmann invariants in radial symmetry scenarios, respectively. Therefore, from Eq. \eqref{2AAA} we interpret the components of ${{\bf\Psi}}$ as ${{\bf\Psi} = (\Psi^{t},\Psi^{r}, \Psi^{\theta}, \Psi^{\varphi})}$ ${= (0,P(r),0,0)}$. Using the divergence theorem in Eq. \eqref{S},

\begin{eqnarray}\label{3AAA}
 S &=& k_{s}\int_{\sigma}{\bf\Psi}\cdot{\bf d\sigma} = k_{s}\int_{V}{\bf\nabla}\cdot{\bf\Psi}dV = k_{s}\int_{V}|s| dV, 
\end{eqnarray}
so, the entropy density is described by  ${s = k_{s}|{\bf\nabla}\cdot{\bf\Psi}|}$.\\

Since the divergence of a vector ${{\bf v}}$ in curved space is given by

\begin{equation}\label{4AAA}
 {\bf\nabla}\cdot{\bf v} = \frac{1}{\sqrt{-g}}\partial_{i}(\sqrt{-g}v^{i}),
\end{equation}
applying this concept to a gravitational entropy density ${s}$ from 
Eq. \eqref{3AAA} and knowing that ${\sqrt{-g} = \sqrt{h}}$, where ${h}$ is the spatial determinant of the metric in the case studied, one has: 

\begin{eqnarray}
 s &=& k_{s}\left|\left(\frac{1}{\sqrt{h}}\frac{\partial}{\partial x^{i}}(\sqrt{h}\Psi^{i})\right)\right| = k_{s}\left|\left(\frac{1}{\sqrt{h}}\frac{\partial}{\partial r}(\sqrt{h}\Psi^{r}) + \frac{1}{\sqrt{h}}\frac{\partial}{\partial \theta}(\sqrt{h}\Psi^{\theta})+ \frac{1}{\sqrt{h}}\frac{\partial}{\partial \varphi}(\sqrt{h}\Psi^{\varphi})\right)\right|.\nonumber\\
 \label{5AAA}
\end{eqnarray}
 Now, knowing that ${{\bf\Psi}}$ has a radial symmetry, consequence already coming from Eq. \eqref{2AAA}, [${{\bf\Psi} = (\Psi^{r}, \Psi^{\theta}, \Psi^{\varphi})}$ ${= (P(r),0,0)}$], we concluded that:

\begin{eqnarray}
 s = k_{s}\left|\left(\frac{1}{\sqrt{h}}\frac{\partial}{\partial r}(\sqrt{h}P(r)) + \frac{1}{\sqrt{h}}\frac{\partial}{\partial \theta}(\sqrt{h}\cdot0)+ \frac{1}{\sqrt{h}}\frac{\partial}{\partial \varphi}(\sqrt{h}\cdot0)\right)\right| = k_{s}\left|\frac{1}{\sqrt{h}}\frac{\partial}{\partial r}(\sqrt{h}P(r))\right|.\nonumber\\
 \label{6AAA}
\end{eqnarray}
So, we see that the spatial components of the vector ${{\bf\Psi}}$, ${\Psi^{i}}$, inserted
in the entropy density equation Eq. \eqref{6AAA} should be ${\Psi^{i} = (P(r),0,0)}$, unlike ${(P(r)/h_{rr},0,0)}$ present in \cite{7}. That is, the only component used is actually ${\Psi^{r} = P(r)}$. The invariant for entropy and gravitational entropy density is guaranteed: ${(\Psi^{\mu}\hat{e}_{\mu})(\Psi_{\nu}\hat{e}^{\nu}) = \Psi^{\mu}\Psi_{\nu}\delta^{\nu}_{\mu} =  \Psi^{i}\Psi_{i} =}$ ${\Psi^{r}\Psi_{r} = P^{2}(r) = P^{2}}$.

Moreover, because the  O.  Rudjord  and O. Gron prescription do not hold for asymmetric metrics, in \cite{8} it is used the form for entropy density in asymmetric metrics like: 

\begin{eqnarray}\label{7AAA}
 s = k_{s}\left|\frac{1}{\sqrt{-g}}\frac{\partial}{\partial r}(\sqrt{-g}P)\right|, 
\end{eqnarray}
where ${\vec{\Psi} = P\hat{e_{r}}}$. Now, you can see that when moving from the case of asymmetric metrics to a diagonal symmetric metric (particular case), the spatial part is written as ${h_{ij} = g_{ij} - g_{i0}g_{j0}/g_{00}}$ and ${\sqrt{-g}\longrightarrow\sqrt{h}}$, recovering
the result of the Eq. \eqref{6AAA}, that was the way we performed our calculations.

\section{Explicit $(W, K)$ terms for wormholes in galactic halo}
\indent
\indent Here we present the calculations of the Weyl and Kretschmann invariants for the metric (\ref{mmhalo}). We use
the software Maple 13 and the program GRTensorII 1,75(R4) \cite{17} as support. 

The Weyl invariant \eqref{IWEYL} for the metric (\ref{mmhalo}) is:
\begin{eqnarray*}
 W_{2}(r) = \frac{1}{12}\frac{1}{r^{6}(r_{s}+r)^{4}}f_{2}(r),
\end{eqnarray*}
with
\begin{eqnarray}\label{awn}
&&f_{2}(r) = \{\bar{r}[r^{2}(l-2)+r_{s}^{2}(l^{2}(rr_{s}^{-1}+1)-5l(rr_{s}^{-1}+1)+6)]+lr^{2}[-l(r+2r_{s}+lr_{s}^{2}r^{-1})\nonumber\\
&&+4r_{s}(r_{s}r^{-1}+2+r_{s}^{-1}r)]+\beta(r)[l\bar{r}((6l^{-1}-5)(r_{s}^{2}-r^{2})+2lrr_{s}+lr^{2}+12l^{-1}-10)\nonumber\\
&&+l^{2}r_{s}^{2}]\}^{2},
\end{eqnarray}
where ${\beta(r) = \ln(1 + r/r_{s})}$, with the numeric values in \cite{1} and \cite{2} $({r_{s} = 6,44.10^{20}m}$, ${r_{th} = \bar{r} = 3,7466.10^{15}m}$ and ${l = 10^{-6})}$. 

From the definition \eqref{IKres} we obtain the Kretschmann invariant of
the metric (\ref{mmhalo}):
\begin{eqnarray*}
 K_{2}(r) = \frac{1}{4r^{6}(r_{s} + r)^{4}}g_{2}(r),
\end{eqnarray*}
with
\begin{eqnarray}\label{akn}
 &&g_{2}(r) =\{ l^{2}\bar{r}r_{s}^{2} [l^{2}((\bar{r}r_{s}(r_{s}(r_{s}+2r)+r))+r_{s}^{2}(2r_{s}r+(14+6r_{s}^{-1}r)r^{2}))]+l(r_{s}(r_{s}^{2}(10r-6\bar{r})\nonumber\\
          &&-r(12\bar{r}r_{s}+4\bar{r}r-28r^{2}))+(2\bar{r}+6r)r^{3})+\bar{r}(r_{s}(r_{s}(17r_{s}+34r)+11r)-6r^{3})-r_{s}(r_{s}(84r\nonumber\\
          &&+28r_{s})r-80r^{3})-(2l^{3}+20l)r^{4}]+\beta(r)(128\bar{r}^{2}r_{s}^{2}r^{2} - 2l^{4}r_{s}^{4}\bar{r}r + 40l^{3}r_{s}r^{4}\bar{r}-12l^{4}r^{3}r_{s}^{2}\bar{r}\nonumber\\
          &&+ 10l^{3}r^{4}_{s}r\bar{r} + 40l^{3}r^{2}r_{s}^{3}\bar{r} + 60l^{3}r^{2}\bar{r}-8l^{4}r_{s}r^{4}\bar{r} - 12l^{2}r_{s}r^{4}\bar{r} -16\bar{r}^{2}r^{4} - 28l^{2}r^{5}\bar{r}-2l^{4}r^{5}\bar{r}\nonumber\\
          &&+ 10l^{3}r^{5}\bar{r} - 168l^{2}r^{3}r_{s}^{2}\bar{r} + 96l^{2}\bar{r}^{2}r_{s}^{2}r^{2}-112l^{2}r^{2}r_{s}^{3}\bar{r} - 28l^{2}rr_{s}^{4}\bar{r} + 48\bar{r}^{2}r_{s}^{4} + 102l^{2}\bar{r}^{2}r_{s}^{4}+6l^{4}\bar{r}^{2}r_{s}^{2}r^{2}\nonumber\\
          &&+ 2l^{3}\bar{r}^{2}r^{4} + 22l^{2}\bar{r}^{2}r^{3}r_{s} - 36l^{3}\bar{r}^{2}r_{s}^{3}r+6l^{4}\bar{r}^{2}r^{3}_{s}r - 34l^{2}\bar{r}^{2}r_{s}^{2}r^{2} - 12l^{3}\bar{r}^{2}r_{s}^{4} - 8r_{s}r^{3}+2l^{4}\bar{r}^{2}r^{3}r_{s}\nonumber\\
          &&+ 34l^{2}\bar{r}^{2}r_{s}^{4} + 144\bar{r}^{2}r_{s}^{3}r + 2l^{4}\bar{r}^{2}r_{s}^{4}-6l^{2}\bar{r}^{2}r^{4})+\beta^{2}(r)(96\bar{r}^{2}r_{s}r^{3} + l^{4}\bar{r}^{2}r_{s}^{4}+24\bar{r}^{2}r^{4}+ 24\bar{r}^{2}r_{s}^{4}\nonumber\\
          &&+ 68l^{2}\bar{r}^{2}r^{3}r_{s}+ 4l^{2}\bar{r}^{2}r+102l^{2}\bar{r}^{2}r_{s}^{2}r^{2}+ 68l^{2}\bar{r}^{2} - 36l^{3}\bar{r}^{2}r_{s}^{2}r^{2}-24l^{3}\bar{r}^{2}r_{s}^{3}r + 17l^{2}\bar{r}^{2}r^{4}+ 6l^{4}\bar{r}^{2}r_{s}^{2}r^{2}\nonumber\\
          &&+ 4l^{4}\bar{r}^{2}r_{s}r^{3}-24l^{5}\bar{r}^{2}r_{s}r^{5} + 17l^{2}\bar{r}^{2}r_{s}^{4}+ 96\bar{r}^{2}r^{3}_{s}r - 6l^{3}\bar{r}^{2}r_{s}^{4}+144\bar{r}^{2}r_{s}^{2}r^{2}+l^{4}-6l^{3}\bar{r}^{2}r^{4})\},
\end{eqnarray}
where ${\beta(r) = \ln(1 + r/r_{s})}$, with the numeric values in \cite{1} and \cite{2} $({r_{s} = 6,44.10^{20}m}$, ${r_{th} = \bar{r} = 3,7466.10^{15}m}$ and ${l = 10^{-6})}$.
The scalar ${P_{2}(r)}$ is given by:
\begin{equation}\label{ap}
 P_{2}(r) = \sqrt{W_{2}(r)/K_{2}(r)} = \sqrt{f_{2}(r)/(3g_{2}(r))}.
\end{equation}


\end{document}